\documentstyle[aps,prb,preprint,graphics]{revtex}

\makeatletter
\def\section{\@startsection {section}{1}{-3.2ex}
{-3.5ex plus -1ex minus -.2ex}{2.3ex plus .2ex}{\bf}}

\def\@biblabel#1{{(#1)}}
\makeatother

\newcounter{themycaption}
\newcommand{\mycaption}[1]{\refstepcounter{themycaption}%
\noindent{\bf Figure \arabic{themycaption}.
}#1\raisebox{0pt}[0pt][4ex]{}\par}

\newcommand{\myfigure}[1]{\newpage\pagestyle{empty}\noindent%
{\bf Figure #1. }A.Ermoshkin, I.Erukhimovich, J.Chen
\raisebox{0pt}[0pt][5ex]{}\\}

\def\eqref#1{(\ref{#1})}

\begin{document}
\normalsize

\title{Statistical Theory of Associating Polymer Solutions: 
Interfacial Properties}

\author{\bf Aleksander V. Ermoshkin$^{1\,{\displaystyle *}}$, Igor
Erukhimovich$^2$, Jeff Z.Y. Chen$^3$}

\address{
$^1$Department of Materials Science and Engineering,
Northwestern University, Evanston, IL, USA 60208-3108 \\
$^2$Department of Physics, Moscow State University,
Moscow, Russia 117234 \\
$^3$Department of Physics, University of Waterloo, Waterloo, Ontario, Canada
N2L 3G1}

\date{November 30, 2001}

\maketitle

\vspace{1cm} \noindent ABSTRACT: The interfacial structure formed
in thermoreversible associating polymer solutions is studied
within the density functional approach based on Flory's arguments
of tree-like configurations of cluster associations. 
The unique characteristics of the interfacial region
can be described in terms of the monomer density along the
interface. For a certain value of the association parameter, which
controls thermoreversible chemical reaction between associating
functional groups, the density profile is not smooth and undergoes
a sudden jump at the point where the number of bonded functional
groups is small. Analytical expression for the interfacial tension is
given and results of numerical calculations are presented.

\newpage

\section{Introduction}
The systems containing molecules capable of forming
thermoreversible chemical bonds with each other are
of significant interest from both fundamental and practical points of view.
These types of systems are often referred to as weak gels \cite{deGennes79}
because the association between molecules under certain physical conditions, in
particular lowering the temperature, leads to the formation
of a thermoreversible infinitely large cluster (network or gel).
At higher temperature, the molecules only form small, separated clusters of finite
size, in the so-called "sol" solution.

The thermodynamical properties of weak gels have been modelled previously
with various theoretical approaches and
approximations.\cite{Coniglio82,Kuchanov88,Coleman92,Tanaka97,Semenov98,Erukh99}
An extensive theoretical study of the sol-gel transition phenomena
was recently presented in Ref. \CITE{Erukh00} where many existing
theoretical models and treatments were surveyed in details,
showing that the pregel regime (or sol fraction) can be well
accounted for with the use of a mean field approach. The original idea
was first  introduced by Flory\cite{Flory41} and
Stockmayer\cite{Stockmayer43} who assumed that all finite-size cluster
associations in the system consist of tree-like
branches only. The analytical extension of this approach to the
postgel regime, where gel fraction appears, gives the classical
theory of gelation now widely accepted.\cite{Rubinstein99} The
sol-gel transition within the frame work of this type of theory
demonstrates smooth transition between the two states in the
thermodynamic behavior, without any singularities typical in terms
of phase transitions.

However, as was first discussed in Ref. \CITE{Erukh79}, it is important to
include the possibility of formation of closed cycles in the model, in order to
describe an infinitely large cluster properly. As the result of such inclusion,
the sol-gel transition has the characteristics of a first-order
thermodynamic phase transition.\cite{Erukh00}

Phase behavior of associating polymer solutions was the target
of these studies and no attention has been paid to their
interfacial properties. In the present paper we study the
interfacial properties related to the region between the two
coexisting phases of associating molecules. As a simple model
system we take the system of $f$-functional monomers, each
carrying $f$ functional groups $A$ that can react with each other
forming reversible chemical bonds $A-A$ (see Fig.
\ref{monomers}a). To obtain the free energy of this model system
we use the classical density functional description which deals
with tree-like clusters only (see Fig. \ref{monomers}b).  The
paper is organized as follows. In Section \ref{FreeEnergy} we give
an explicit expression for the free energy functional of
inhomogeneous solution of tree-like associating molecules. In
Section \ref{SurfTension} we describe the analytical and numerical
procedures for minimization of the free energy functional and
derive the expression for the interfacial tension. Readers who are
not interested in technical details might skip sections
\ref{FreeEnergy} and \ref{SurfTension}, and go to section
\ref{Results} directly where we discuss the obtained results.

\section{Free energy of an inhomogeneous system within the functional density
approach} \label{FreeEnergy} Consider a solution of $f$-functional
monomers distributed in space with a nonuniform monomer density
$\rho({\bf r})$. The free energy of this system can be divided
into two parts\cite{Lifshitz69}
\begin{equation}
F([\rho({\bf r})],T)=F^*([\rho({\bf r})],T)+F_{str}([\rho({\bf r})],T),
\label{Frho}
\end{equation}
where the first term on the right-hand side contains the
contribution from the excluded-volume interaction between monomer
units and the second term is related to the structural free energy
of the system of an ideal solvent. The excluded-volume
interaction, in particular, can be accounted for by the following
expression, which was proposed by Flory\cite{Flory53} based on a
lattice model consideration,
\begin{equation}
F^*([\rho({\bf r})],T)=\frac{T}{v}\int\Bigl\{[1-v\rho({\bf r})]\ln[1-v\rho({\bf
r})]-\chi v^2\rho^2({\bf r})\Bigr\}d{\bf r},
\label{Fsrho}
\end{equation}
where $T$ is the temperature of the system in $k_B$ units, $v$ is the volume of one lattice cell
(the volume of one monomer) and $\chi$ is the Flory-Huggins constant, with $\chi T$
being the energy gain per monomer-monomer contact.

Our main purpose in this section is to introduce the structural term 
$F_{str}([\rho({\bf r})],T)$ as derived within the density functional approach 
in Ref. \CITE{Erukh79,Erukh95}.
Assuming that the average distance $r_0$ between monomers is much less then
the average length $a$ of a chemical bond $A-A$,
\begin{equation}
r_0\ll a,
\label{r0lla}
\end{equation}
we write the structural free energy in the form
\begin{equation}
F_{str}([\rho({\bf r})],T)/T=
-S_{id}[\rho({\bf r})]+{\rm min}\Bigl\{S_{comb}[\rho({\bf r}),\rho_A({\bf r})]
-S_A[\rho_A({\bf r})]\Bigr\}.
\label{Fstrrho}
\end{equation}
Here the first term
\begin{equation}
S_{id}[\rho({\bf r})]=-\int\rho({\bf r})\ln\left(\frac{v\rho({\bf r})}{f!}\right)
d{\bf r}
\label{Sid}
\end{equation}
corresponds to the translational entropy of the monomer units
in space. The minimum of the second term
is considered in the configurational space of all possible density
distributions $\rho_A({\bf r})$. We also have
\begin{equation}
S_{comb}[\rho({\bf r}),\rho_A({\bf r})]=-\int f\rho({\bf r})
\left[\Gamma({\bf r})\ln(\Gamma({\bf r}) +
(1-\Gamma({\bf r}))\ln(1-\Gamma({\bf r}))\right]
d{\bf r},
\label{Scomb}
\end{equation}
where $\Gamma({\bf r})\equiv\rho_A({\bf r})/f\rho({\bf r})$ is the
local content of conversion, corresponding to the entropy of
selecting reacted functional groups distributed with the density
$\rho_A({\bf r})$.

The other term in eq. \eqref{Fstrrho}, $S_A$, corresponds to the entropy of
thermoreversible chemical bonds. Here we assume that under condition
\eqref{r0lla} $S_A$ depends only on the density
of reacted functional groups and it does not depend on a particular monomer 
structure.
Thus $S_A$ can be calculated for the simplest  system of fully reacted monofunctional units.
From Appendix \ref{SA_calc} we have:

\begin{equation}
S_A[\rho_A({\bf r})]=\int\frac{\rho_A(\bf r)}{2}\ln\left[
\frac{\rho_A({\bf r})(\hat g\psi)({\bf r})}{e\psi({\bf r})}\right]d{\bf r},
\label{SA11}
\end{equation}
where $\rho_A({\bf r})$ and $\psi({\bf r})$ are dependent 
equation
\begin{equation}
\rho_A({\bf r})=\psi({\bf r})(\hat g\psi)({\bf r})=\psi({\bf r})\int g({\bf r}-{\bf
r'})\psi({\bf r'})d{\bf r}
\label{rhoApsihatg}
\end{equation}
with $g({\bf r})$ being the statistical weight of a bond with length $\bf r$.
For a homogeneous system we can rewrite eq. \eqref{SA11} in the form
\begin{equation}
S_A(V,\rho_A)=V\frac{\rho_A}{2}\ln\frac{k\rho_A}{e}
\label{SA2}
\end{equation}
where
\begin{equation}
k=\int g({\bf r})d{\bf r}.
\label{g}
\end{equation}
The association caused contribution \eqref{SA2} was also obtained by Semenov and 
Rubinstein in Ref. \CITE{Semenov98} who considered the homogeneous systems only.
After minimization of the second term in eq. \eqref{Fstrrho} we
recover the expression which is widely used to consider phase
behavior of associating polymer systems
\begin{equation}
F_{str}^{hom}(T,V,\rho)/(TV)=\rho\ln\left(\frac{f!v\rho}{e}\right)+
f\rho\left(\frac{\Gamma}{2}+\ln(1-\Gamma)\right),
\label{Fstrhom}
\end{equation}
and
\begin{equation}
fk\rho=\frac{\Gamma}{(1-\Gamma)^2}.
\label{massaction}
\end{equation}
We shall not discuss details of phase behavior in this paper, but
would like to mention that for certain values of parameters $\chi$
and $g$ it is more preferable for the system to be separated into
two macroscopic sol phases with respectively different densities
of monomers $\rho_1$ and $\rho_2$ which could be obtained from the
equations
\begin{eqnarray}
\mu_1(T,\rho_1)&=&\mu_2(T,\rho_2) \label{mupi1} \\
\pi_1(T,\rho_1)&=&\pi_2(T,\rho_2) \label{mupi2}
\end{eqnarray}
where $\mu_i$ and $\pi_i$ are the chemical potential and osmotic
pressure of phase $i$ ($i=1,2$). We should also point out that if
the density of monomers in the phase exceeds the critical value
$\rho^*$
\begin{equation}
fk\rho^*=\frac{f-1}{(f-2)^2}
\label{rhostar}
\end{equation}
then the gel phase (infinitely large network) will be formed.

In the next section we consider only the systems where two sol phases with
different densities are present,  describe the procedure that minimizes
free energy functional \eqref{Frho}, and derive the expression for the
interfacial tension between the coexisting phases.

\section{Minimization of the free energy functional and the interfacial
tension}
\label{SurfTension}
In order to proceed further, we introduce a coordinate system so that the
$z=0$ surface coincides with the flat interface separating the
two sol phases. For a density variation whose typical
correlation length is much greater than the basic Kuhn length
associated with a single bond, the operator $\hat g$ can be 
approximated \cite{Grosberg94}
\begin{equation}
\hat g=k\left(1+\frac{a^2}{6}\frac{d^2}{dz^2}\right),
\label{hatg}
\end{equation}
where the chemical equilibrium constant $k$ is defined in eq. \eqref{g} and
$a$ is the average length of a chemical bond. The free energy per unit surface can then
be expressed as
\begin{equation}
\frac{F}{S}=\frac{T}{v}\,{\rm min}\left\{\,\int\limits_{-L_1}^{L_2}
{\cal F}\biggl(\phi(z),\phi_A(z),\dot\phi_A(z)\biggr)dz\right\}
\label{Fint}
\end{equation}
where $S$ is the area of the interface, $L_1$ and $L_2$ are the linear dimensions of each
phase normal to the interface, $\phi=v\rho$ is the volume fraction of monomers
and we will refer to $\phi_A=v\rho_A$ as the volume fraction of reacted functional groups.
The dot symbol represents the derivative with respect to $z$
\begin{equation}
\dot\phi_A\equiv\frac{d\phi_A}{dz}.
\label{dotrhoAdef}
\end{equation}
The density free energy ${\cal F}$, which is a functional of the local volume fractions of 
monomers and reacted  functional groups, takes the form
\begin{equation}
{\cal F}(\phi,\phi_A,\dot\phi_A)={\cal F}_{bls}(\phi)+{\cal F}_{comb}(\phi,\phi_A)+{\cal
F}_A(\phi_A,\dot\phi_A).
\label{calF}
\end{equation}
The first term in (\ref{calF}) corresponds to the free energy of the broken-link system
\begin{equation}
{\cal F}_{bls}(\phi)=\phi\ln\phi+(1-\phi)\ln(1-\phi)-\chi\phi^2,
\label{calFbls}
\end{equation}
the second term in (\ref{calF}) is simply combinatorial
\begin{equation}
{\cal F}_{comb}(\phi,\phi_A)=f\phi\biggl(\Gamma\ln\Gamma+(1-\Gamma)\ln(1-\Gamma)\biggr), \quad
\Gamma=\frac{\phi_A}{f\phi},
\label{calFcomb}
\end{equation}
and the third one corresponds to the entropy of the chemical bonds
\begin{equation}
{\cal F}_A(\phi_A,\dot\phi_A)=-\frac{\phi_A}{2}\ln\frac{g\phi_A}{e}+\frac{a^2}{48}\frac{\dot\phi_A^2}{\phi_A}
\label{calFAhatg}
\end{equation}
where we have introduced a dimensionless chemical reaction constant
\begin{equation}
g=k/v.
\label{gkv}
\end{equation}

While the free energy across the interface is given by eq. \eqref{Fint}, the surface
tension is defined through considering the excess free energy in reference to the
bulk phases. We can then write
\begin{equation}
\sigma=\frac{T}{v}\,\,{\rm min}\left\{\,\int\limits_{-L_1}^{0}\left({\cal
F}(\phi,\phi_A,\dot\phi_A)-{\cal
F}_1^{hom}\right)dz
+\int\limits_{0}^{L_2}\left({\cal F}(\phi,\phi_A,\dot\phi_A)-{\cal F}_2^{hom}\right)dz\right\}
\label{sigma}
\end{equation}
where the terms
\begin{eqnarray}
{\cal F}_i^{hom}\equiv{\cal F}^{hom}(\phi_i)={\cal F}_{sbl}(\phi_i)+
f\phi_i\left(\frac{\Gamma_i}{2}+\ln(1-\Gamma_i)\right) && \nonumber\\
fg\phi_i=\frac{\Gamma_i}{(1-\Gamma_i)^2}&&\quad(i=1,2),
\label{Fhomi}
\end{eqnarray}
are simply free energies of two homogenous phases with monomer
volume fractions $\phi_1$ and $\phi_2$ respectively on the left
and right sides of the interface.

The minimization of the functional \eqref{sigma} is considered with the constraint
\begin{equation}
\int\limits_{-L_1}^{0}(\phi(z)-\phi_1)dz+\int\limits_{0}^{L_2}(\phi(z)-\phi_2)dz=0
\label{Nphi1}
\end{equation}
and the volume fractions $\phi_1$ and $\phi_2$ are obtained from
eqs. \eqref{mupi1}, \eqref{mupi2}, or, written more explicitly,
\begin{eqnarray}
\left.\frac{\partial{\cal F}^{hom}}{\partial\phi}\right|_{\phi=\phi_1}&=&
\left.\frac{\partial{\cal F}^{hom}}{\partial\phi}\right|_{\phi=\phi_2} \\
(\phi_1-\phi_2)\left.\frac{\partial{\cal F}^{hom}}{\partial\phi}\right|_{\phi=\phi_1}
&=&{\cal F}^{hom}_1-{\cal F}^{hom}_2
\label{phi1phi2}
\end{eqnarray}

Since any contributions to the free energy density which are
independent of $\phi$, or linear in $\phi$, could be dropped or
added,\cite{deGennes79} we can introduce a new free energy density
function $\widetilde{\cal F}(\phi,\phi_A,\dot\phi_A)$,
\begin{equation}
\widetilde{\cal F}(\phi,\phi_A,\dot\phi_A)={\cal F}(\phi,\phi_A,\dot\phi_A)+A\phi+B
\label{wcalF}
\end{equation}
where the parameters $A$ and $B$ are to be determined from equations
\begin{eqnarray}
{\cal F}^{hom}_1+A\phi_1+B&=&0 \label{AB1} \\
{\cal F}^{hom}_2+A\phi_2+B&=&0 \label{AB2}
\end{eqnarray}
We can now rewrite eq. \eqref{sigma} in the form
\begin{equation}
\sigma=\frac{T}{v}\,\,{\rm min}\left\{\,\int\limits_{-\infty}^{+\infty}\widetilde{\cal F}
(\phi,\phi_A,\dot\phi_A)dz\right\}
\label{sigma1}
\end{equation}
where we have rewritten the integration limits in terms of infinity since $\widetilde{\cal F}$ 
approaches
zero far from the interface.

According to a standard minimization procedure we introduce two Euler's equations
\begin{eqnarray}
\frac{\partial\widetilde{\cal F}}{\partial\phi}&=&0 \label{E1} \\
\frac{\partial\widetilde{\cal F}}{\partial\phi_A}-\frac{d}{dz}\frac{\partial\widetilde
{\cal F}}{\partial\dot\phi_A}&=&0
\label{E2}
\end{eqnarray}
As the function $\widetilde{\cal F}$ is independent of $z$, a first integral can be used to replace
eq. \eqref{E2}:
\begin{equation}
\widetilde{\cal F}-\dot\phi_A\frac{\partial\widetilde{\cal F}}{\partial\dot\phi_A}=0.
\label{E2int}
\end{equation}
From eq. \eqref{E1}
\begin{equation}
\frac{\partial{\cal F}_{bls}}{\partial\phi}+A+f\ln(1-\Gamma)=0
\label{rhoGamma}
\end{equation}
one can find the equilibrium profile of conversion $\Gamma$

\begin{equation}
\Gamma(\phi)=1-\exp\left(-\frac{1}{f}\frac{\partial{\cal F}_{bls}}{\partial\phi}-\frac{A}{f}\right).
\label{Gammarho}
\end{equation}
This expression is valid across the entire interface and it gives a relation between
conversion $\Gamma$ and monomer volume fraction $\phi$.

From eq. \eqref{E2int} it follows that
\begin{equation}
\frac{a^2}{48}\frac{\dot\phi_A^2}{\phi_A}={\cal F}_{bls}+A\phi+B+
f\phi(\Gamma\ln\Gamma+(1-\Gamma)\ln(1-\Gamma))-\frac{\phi_A}{2}\ln\frac{g\phi_A}{e}
\label{dotrhoArhoA}
\end{equation}
Taking into account eq. (\ref{dotrhoArhoA}) and the definition of $\Gamma$ (see eq. \eqref{calFcomb})
we can now rewrite eq. \eqref{sigma} for the interfacial tension
\begin{equation}
\sigma=\frac{2T}{v}~\int\limits_{-\infty}^{+\infty}{\cal F}_\sigma(\phi(z))\,dz,
\label{sigmaFsigma} 
\end{equation}
where
\begin{equation}
{\cal F}_\sigma(\phi)={\cal F}_{bls}(\phi)+A\phi+B+
f\phi\left(\frac{\Gamma(\phi)}{2}\ln\frac{e\Gamma(\phi)}{fg\phi(1-\Gamma(\phi))^2}+
\ln(1-\Gamma(\phi))\right)
\label{Fsigma}
\end{equation}
Furthermore, $\dot\phi(z)$ can be written as
\begin{equation}
\dot\phi_A=f(\dot\phi\Gamma+\phi\dot\Gamma)=
f\dot\phi\left(\Gamma+\phi\frac{\partial\Gamma}{\partial\phi}\right)
\label{dotrhoA}
\end{equation}
Finally, with the help of eqs. \eqref{dotrhoArhoA} and \eqref{Fsigma}, $\phi(z)$ obeys a simple
differential relation
\begin{equation}
\dot\phi^2=\frac{48}{a^2}\frac{\phi\Gamma{\cal F}_\sigma}
{f\left(\Gamma+\phi\frac{\partial\Gamma}{\partial\phi}\right)^2}
\label{dotrho}
\end{equation}
where
\begin{equation}
\frac{\partial\Gamma}{\partial\phi}=\frac{1}{f}\frac{\partial^2{\cal
F}_{bls}}{\partial\phi^2}(1-\Gamma).
\label{partialGamma}
\end{equation}

Therefore, the volume fraction profile $\phi(z)$ can be obtained using the simple iterative numerical
procedure
\begin{equation}
\phi(z_{i+1})=\phi(z_i)+\dot\phi(\phi(z_i))(z_{i+1}-z_i)
\label{iter}
\end{equation}
where $\dot\phi(\phi)$ is defined by eq. (\ref{dotrho}). The interfacial tension $\sigma$ can then be
easily found through eq. (\ref{sigmaFsigma}) after the calculation of the density.

\section{Results and discussions}
\label{Results}
In this paper we study the temperature dependence of the interfacial tension in the solution of
associating monomers capable of forming thermoreversible chemical bonds with each other.
The functionality $f$ of a monomer unit defines the structure of associations that could be formed .
If $f=2$ then the system would contain linear molecules of various lengths. For $f=3$ more complex
structures could be formed as shown in Figure \ref{monomers}.

The temperature dependence of the Flory parameter $\chi$ is considered to have the simple form
\begin{equation}
\chi=\frac{\theta}{2T},
\label{chi}
\end{equation}
although more exotic temperature dependence could be introduced. We also introduce
parameter $g$ defined through eq. \eqref{gkv} in the form
\begin{equation}
g=g_0\exp(-E\chi),
\label{gEchi}
\end{equation}
where $g_0$ is the chemical reaction constant at high temperatures and $E\theta/2$ is the energy of a
chemical bond formed between two functional groups.

As the first step, we consider the interfacial tension in a system of bi-functional units ($f=2$).
If the chemical reaction constant $g$ is high enough then all functional groups would be reacted and
the system would contain one infinitely large linear chain. In this limiting case our theory should
recover the treatment for the interfacial tension proposed by Helfand and Tagami\cite{Helfand71a,Helfand71b}.
Writing free energy of the solution containing infinitely long linear chain in the form\cite{Ermoshkin96}
\begin{equation}
F=\frac{T}{v}\int\left[\frac{a^2}{24}\frac{(\nabla\phi)^2}{\phi}+(1-\phi)\ln(1-\phi)+\chi\phi(1-\phi)
\right]d{\bf r}
\label{FHelfand}
\end{equation}
one could obtain the expression for the interfacial tension $\sigma_0$ corresponding to the case
where the rich phase (containing all chain monomers) and poor phase (only solvent molecules)
are independently formed
\begin{equation}
\sigma_0=\frac{Ta}{v\sqrt{6}}\int\limits_{0}^{\phi_0}
\left(\frac{1-\phi}{\phi}\ln(1-\phi)+\chi(1-\phi)-\mu_0\right)^{1/2}d\phi
\label{sigma_0}
\end{equation}
Here $T\mu_0$ is the chemical potential and $\phi_0$ is the volume fraction of monomers in the rich
phase. These two parameters could be found from the equations
\begin{eqnarray}
\mu_0&=&\chi(1-2\phi_0)-\ln(1-\phi_0)-1 \\
\mu_0\phi_0&=&\chi\phi_0(1-\phi_0)+(1-\phi_0)\ln(1-\phi_0)
\end{eqnarray}
which balance chemical potentials and osmotic pressures of bi-functional monomer units in the poor and
rich phases.

Figure \ref{sigma_linear} shows the temperature dependence of interfacial tension $\sigma_0$ given by
eq. \eqref{sigma_0} (open circles) together with $\sigma$ based on
eq. \eqref{sigmaFsigma} for $E=0$ and different values of interaction parameters $g_0$ introduced
in eq. \eqref{gEchi}. For large $g_0$ ($\ln g_0=20$, dotted line) we see that
our result agrees exactly with the Helfand-Tagami theory thus confirming the model in the ($f=2$,
$g\gg1$) limit.

Our model also accounts for the situation where a polydispersed system of linear chains system can be
formed for smaller $g_0$. For smaller values of $g_0$ ($\ln g_0=2$, dashed line, and $\ln g_0=0$, solid
line) when the reaction of functional groups is incomplete one can see the singularities that might 
occur in the $\sigma(T)$ behavior. Distinct singularity
points can be identified where sudden changes in curvatures occur.

Figure \ref{int_ten} demonstrates the interfacial tension of the $f=3$ case where
the branched associates may coexist across the interface and one can also observe the similar
singularity behavior for $\sigma(T)$.

The described types of singularities in the interfacial tension
behavior arise due to the singularities that can occur in the
behavior of the density profiles $\phi(z)$ and $\phi_A(z)$ along
the interface (smooth profiles without any singularities are shown
at the top of Fig. \ref{smooth}). For a certain value of parameters
$\chi$ and $g$ one can expect a sudden change in the slope of
$\phi_A(z)$ as shown in Fig. \ref{slope}. This picture is observed
if $\Gamma+\phi\,\partial\Gamma/\partial\phi$ introduced in eq.
(\ref{dotrho}) becomes negative in the interval ($\phi'$,$\phi''$)
as shown at the bottom of Fig. \ref{slope}. According to eq.
\eqref{dotrhoA} the slope of $\phi(z)$ becomes infinite at $z'$
and $z''$ which correspond to $\phi'$ and $\phi''$ and the slope
of $\phi_A(z)$ is negative in the entire interval $(z',z'')$.

The more complex behavior is observed when the equilibrium
conversion  $\Gamma(\phi)$ defined by eq. (\ref{Gammarho}) becomes
negative, as shown at the bottom of Fig. \ref{jump} for the
interval $(\phi'',\phi''')$. Physically  $\Gamma$ is a positive
quantity and therefore the region $(\phi'',\phi''')$ should be
skipped when calculating $\sigma$ defined by \eqref{sigmaFsigma}.
Skipping this $(\phi'',\phi''')$ interval leads to the sudden jump
of the volume fraction profile $\phi(z)$ at the point $z''$ as
shown at the top of  Fig. \ref{jump}. Similar noncontinuous
behavior of polymer density was previously discussed in Ref.
\CITE{Lifshitz74} where the authors considered a polymer globule
formed by a long polymer chain with saturated physical bonds. This
type of behavior could be explained as follows. As one goes across
the interface (see the top of Fig. \ref{jump}) the volume fraction
$\phi$ increases until it reaches the value $\phi''$ at the point
$z''$. At the same time  the number of reacted functional groups
(or reduced density of reacted functional groups $\rho_A$)
decreases and becomes infinitely small as  $\phi$ reaches $\phi''$
(at this point conversion $\Gamma$ is zero). The fact that
$\phi_A=0$ ensures that there would be no loss of conformational
entropy for the system associated with the jump of the volume
fraction at the point $z''$ because square gradient term (see eq.
\eqref{calFAhatg}) depends only on derivative of $\phi_A$ which is
continuous at this point.

Hence, the behavior of interfacial profile $\sigma(T)$ shown in
Figure \ref{int_ten} could be explained as follows. As the
temperatures of the system decreases and reaches $T_1$ the
interface first appears and the interfacial tension starts to
increase. As temperature reaches the value of $T_2$ the change in
the curvature of $\sigma(T)$ profile is observed. This curvature
change is connected to the appearance of the negative slope in the
behavior of $\phi_A(z)$ (as shown in Fig. \ref{slope}). When the
temperature of the system reaches $T_3$ one observes the change in
the slope of $\sigma(T)$ profile due to the discontinuity of the
monomer volume fraction $\phi(z)$ (see Fig. \ref{jump}) which
first appears at this temperature.

\section{Conclusion}
In this paper, we analyze the interfacial structure and calculate
the temperature dependence of the interfacial tension $\sigma$ in
associating solutions of $f$-functional monomers. We find that
there are two types of singularities which occur in the behavior
of the density profiles along the interface. At certain
temperatures the density of reacted functional groups may decrease
within some region inside the interface while the polymer density
always increases. Also, the polymer density may exhibit a sudden
jump at the point where the density of reacted functional groups
becomes infinitely small. The first of these singularities leads
to the change in the curvature of  the $\sigma(T)$ profile and the
second one changes the slope of that profile.

\section*{Acknowledgments}
This work was supported by the NATO Science Fellowship awarded to
A. Ermoshkin, the INTAS Grant No. 99-01852 awarded to I. Erukhimovich,
and a NATO collaborative Research Grant and an NSERC grant awarded 
to J. Z. Y. Chen.

\appendix

\section{Calculation of the entropy of thermoreversible chemical 
bonds\cite{Erukh79,Erukh95}}
\label{SA_calc}

In this Appendix we cosider the system of fully reacted monofunctional 
monomers units. If $\rho_A({\bf r})$ is the monomer density, then the entropy 
of chemical bonds $S_A$ is defined as follows
\begin{equation}
S_A[\rho_A({\bf r})]=S_{pair}[\rho_A({\bf r})]-
S_{id}[\rho_A({\bf r})].
\label{SA}
\end{equation}
where $S_{pair}$ is the entropy of dimers formed due to the associations and $S_{id}$
defined by eq. \eqref{Sid} corresponding to the translational entropy of monomers.
We assume that external field $\varphi({\bf r})$ produces an equilibrium
density  $\rho_A({\bf r})$, and in the mean field approximation
\begin{equation}
TS_{pair}[\rho_A({\bf r})]=
E[\varphi({\bf r})]-F_{pair}[\varphi({\bf r})]
\label{Spair}
\end{equation}
Here $E[\varphi(\bf r)]$ and $F_{pair}[\varphi(\bf r)]$ are the energy and free
energy of dimers in the field $\varphi({\bf r})$ respectively. They
satisfy the following expressions\cite{Grosberg94}
\begin{equation}
E[\varphi({\bf r})]=\int \rho_A({\bf r})\varphi({\bf r})d{\bf r}
\label{Evarphi}
\end{equation}
\begin{equation}
F_{pair}[\varphi({\bf r})]=-T\ln\left[
\frac{\biggl({\displaystyle\int} \psi({\bf r})(\hat g\psi)({\bf r})d{\bf r}\biggr)^N}
{2^N N!\,v^{2N}}\right]
\label{Fpair}
\end{equation}
where $N$ is the total number of dimers
\begin{equation}
N=\frac{1}{2}\int\rho_A({\bf r})d{\bf r}
\label{N}
\end{equation}
and
\begin{eqnarray}
\psi({\bf r})&=&\exp\left(-\frac{\varphi({\bf r})}{T}\right) \label{psiphi} \\
(\hat g\psi)({\bf r})&=&\int g({\bf r}-{\bf r'})\psi({\bf r'})d{\bf r}
\label{psihatg}
\end{eqnarray}
with $g({\bf r})$ corresponding to the statistical weight of a bond with length $\bf r$.
According to general thermodynamic relations\cite{Landau80} the density of functional groups can
be written  as
\begin{eqnarray}
\rho_A({\bf r})=\frac{\delta F_{pair}[\varphi({\bf r})]}{\delta \varphi({\bf r})}
=2N\frac{\psi({\bf r})(\hat g\psi)({\bf r})}{{\displaystyle\int} \psi({\bf r})(\hat g\psi)({\bf r})}
\label{rhoAF2}
\end{eqnarray}
Using eqs. \eqref{N} and \eqref{rhoAF2} one can rewrite eq. \eqref{Fpair} in the form
\begin{equation}
F_{pair}[\varphi({\bf r})]=-T\int\frac{\rho_A(\bf r)}{2}\ln\left[
\frac{\psi({\bf r})(\hat g\psi)({\bf r})}{\rho_A({\bf r})}
\frac{e}{v^2}\right]d{\bf r}
\label{Fpair1}
\end{equation}
Eliminating $\varphi({\bf r})$ from eq. \eqref{Evarphi} with the help of eq. \eqref{psiphi}
and substituting the result into eq. \eqref{Spair} together with expression \eqref{Fpair1}
we obtain
\begin{equation}
S_{pair}[\rho_A({\bf r})]=\int\frac{\rho_A(\bf r)}{2}\ln\left[
\frac{(\hat g\psi)({\bf r})}{\psi({\bf r})\rho_A({\bf r})}
\frac{e}{v^2}\right]d{\bf r},
\label{Spair1}
\end{equation}
and for the entropy of chemical bonds given by eq. \eqref{SA} we finally get
\begin{equation}
S_A[\rho_A({\bf r})]=\int\frac{\rho_A(\bf r)}{2}\ln\left[
\frac{\rho_A({\bf r})(\hat g\psi)({\bf r})}{e\psi({\bf r})}\right]d{\bf r},
\label{SA1}
\end{equation}
which is the formula used in eq. \eqref{SA11}.


\newpage

\section*{Figure captions}

\mycaption{Formation of a reversible chemical bond $A-A$ between two
three-functional monomers (a) and an example of a tree-like cluster formed by
three-functional monomers (b).}
\label{monomers}

\mycaption{Interfacial tension $\sigma$ given by eq. \eqref{sigmaFsigma}
($v=1$, $a=1$) as a function of temperature for bi-functional units with $f=2$,
$E=0$, and different values of $\ln g_0$. Open circles correspond
to the limited case given by eq. \eqref{sigma_0}. }
\label{sigma_linear}

\mycaption{Interfacial tension $\sigma$ given by eq. \eqref{sigmaFsigma}
($v=1$, $a=1$) as a function of temperature for three-functional units with $f=3$ and
values of interaction parameters shown in the upper right corner. }
\label{int_ten}

\mycaption{Equilibrium profiles of volume fraction $\phi$ (solid line)
and volume fraction of reacted functional groups $\phi_A$ (dash line)
along the interface in  the case of $f=3$ and values of interaction
parameters shown in the upper right corner (top). Also shown are the 
equilibrium conversion $\Gamma$ 
(solid line) and $\Gamma+\phi\,\partial\Gamma/\partial\phi$
(dash line) as functions of volume fraction $\phi$.}
\label{smooth}

\mycaption{See caption for Fig. \ref{smooth}.}
\label{slope}

\mycaption{See caption for Fig. \ref{smooth}.}
\label{jump}


\myfigure{\ref{monomers}}
\includegraphics{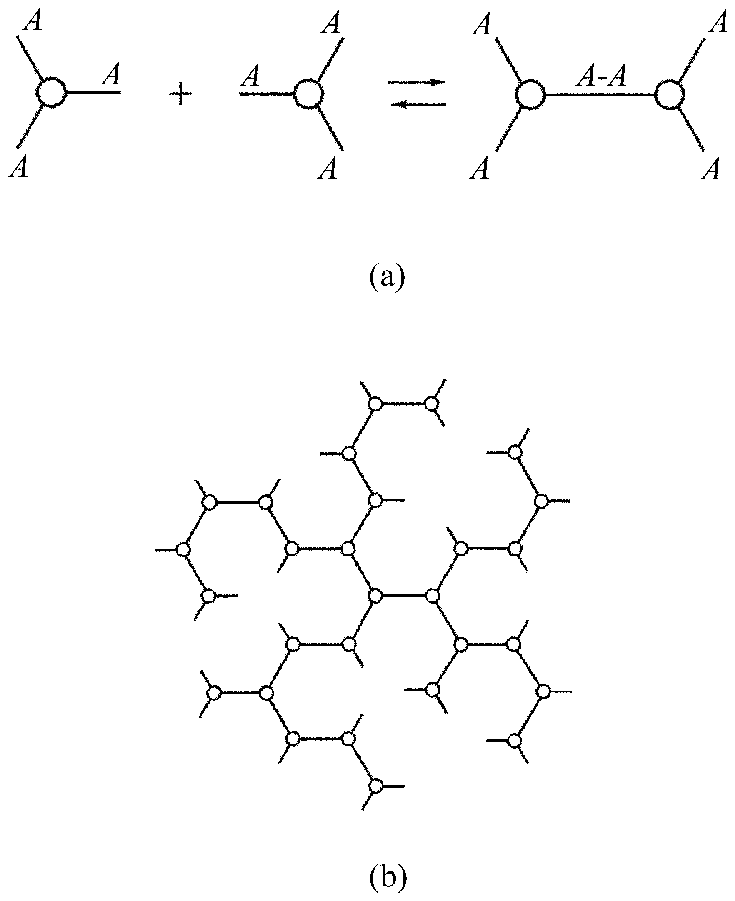}

\myfigure{\ref{sigma_linear}}
\includegraphics{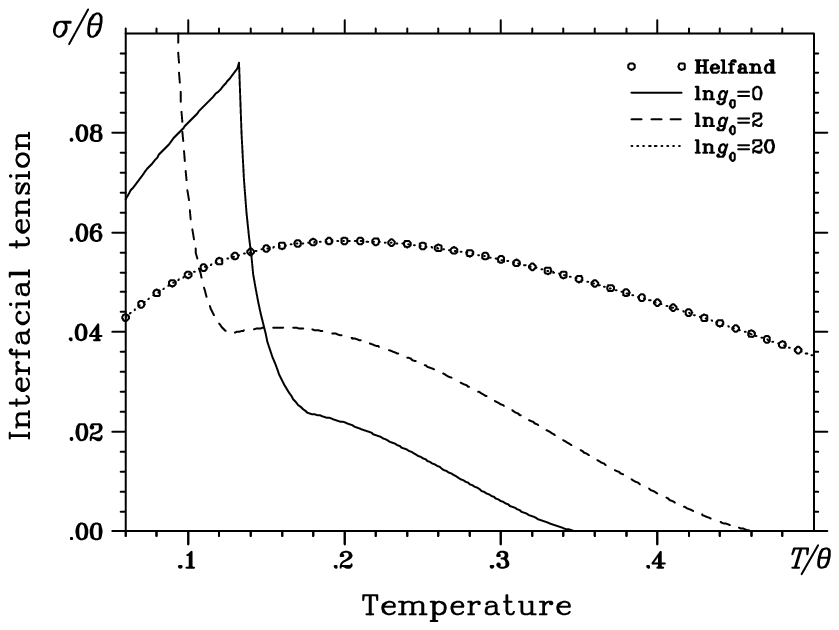}

\myfigure{\ref{int_ten}}
\includegraphics{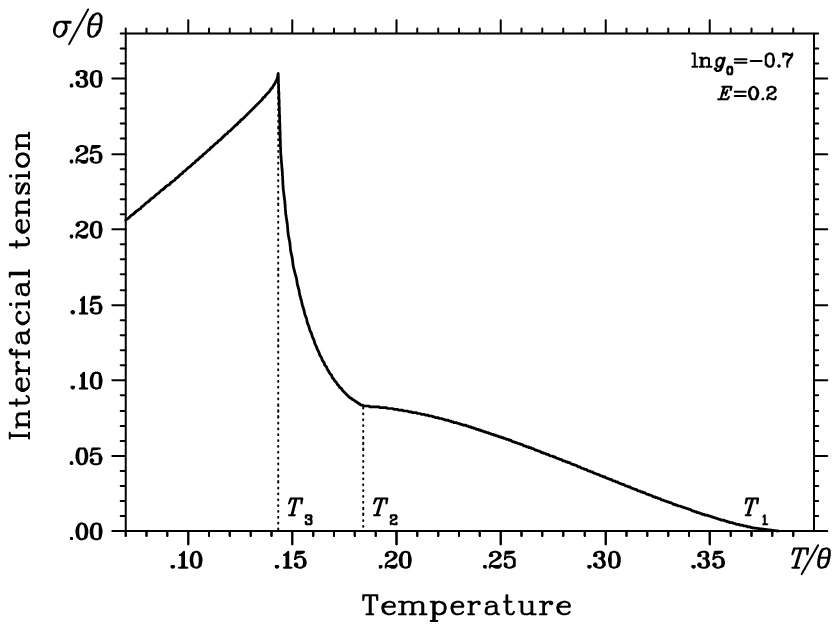}

\myfigure{\ref{smooth}}
\includegraphics{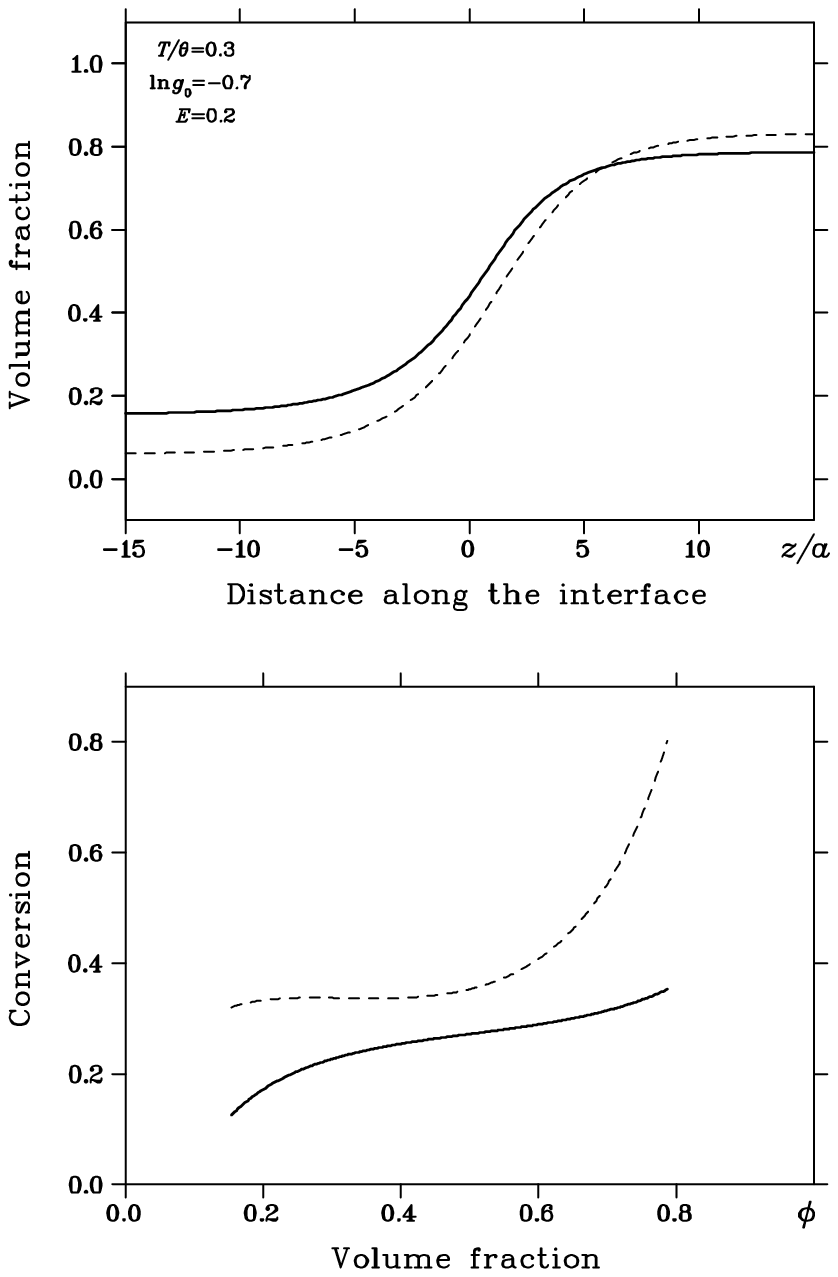}

\myfigure{\ref{slope}}
\includegraphics{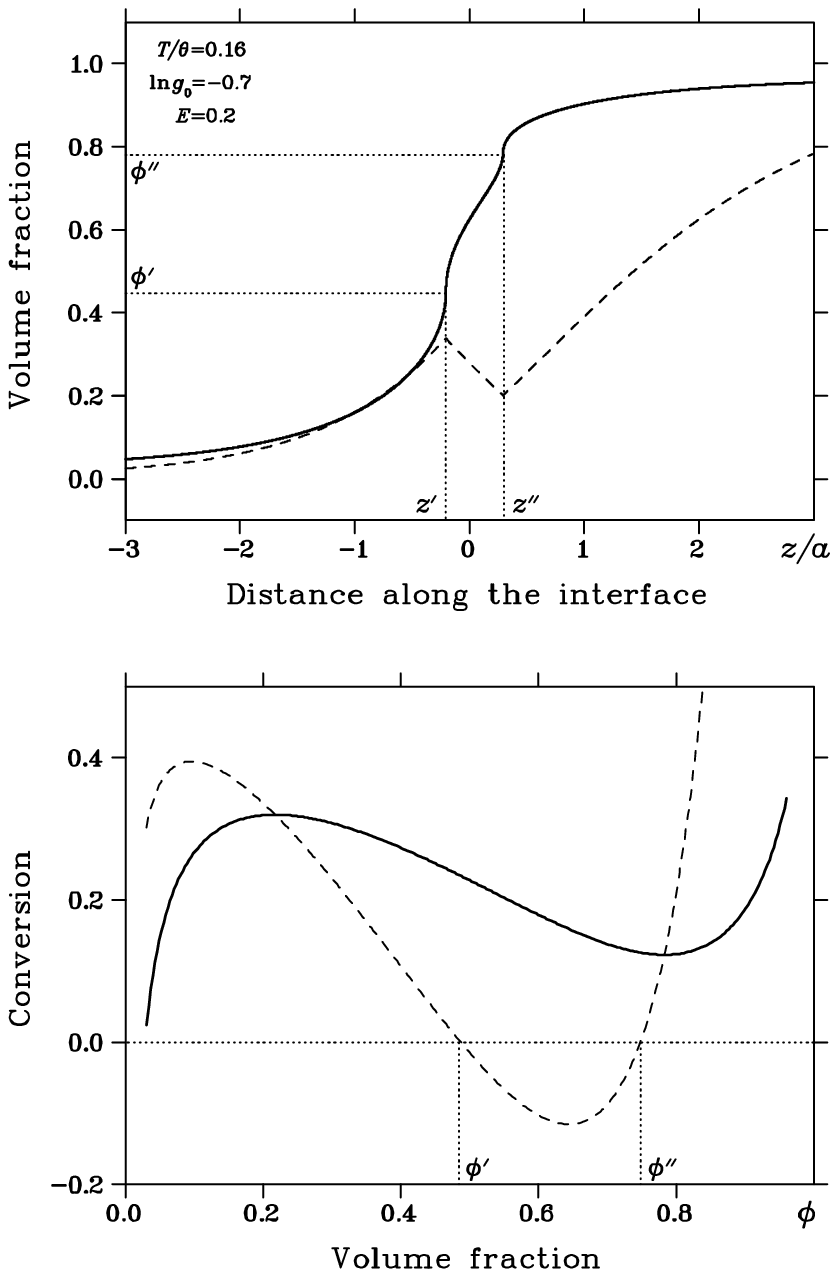}

\myfigure{\ref{jump}}
\includegraphics{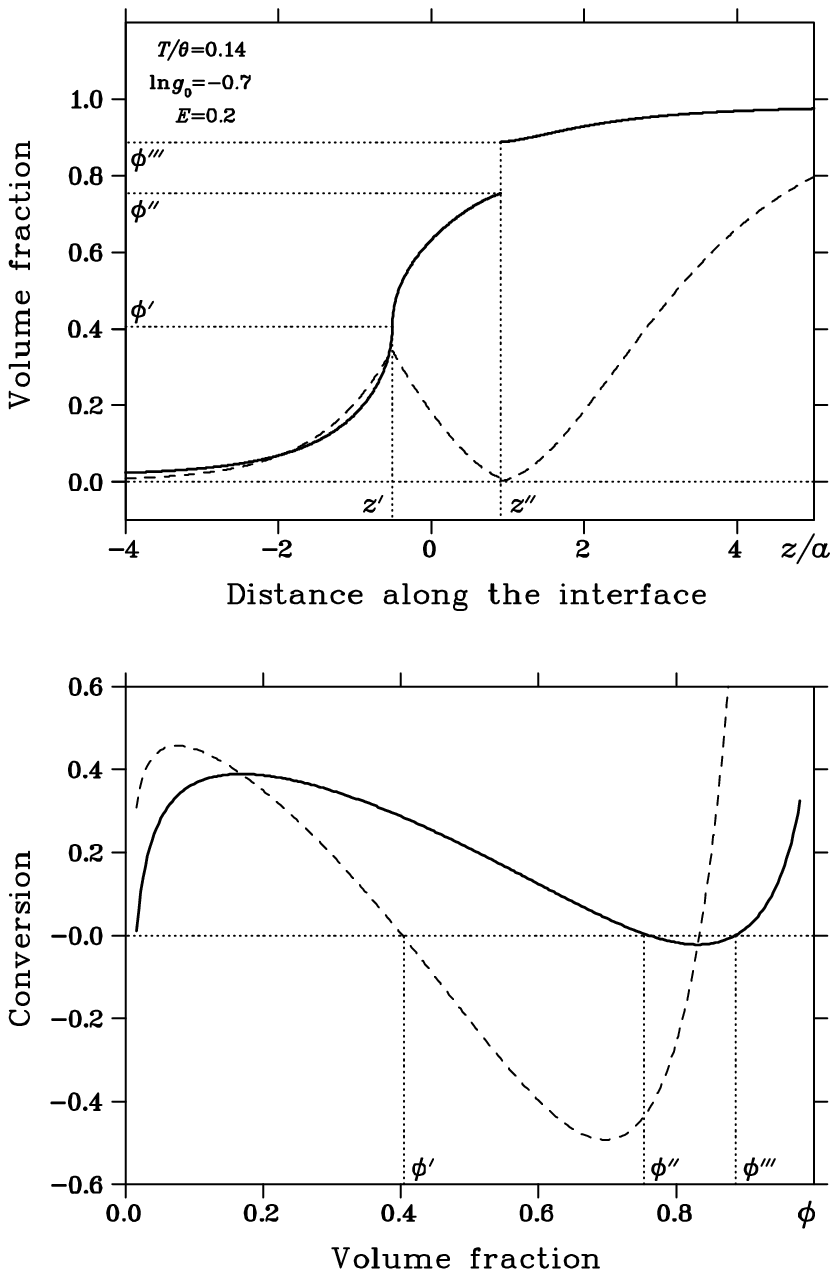}

\end{document}